\documentclass{PoS}

\title{Solution to new sign problems with Hamiltonian Lattice Fermions}

\ShortTitle{Solution to new sign problems with Hamiltonian Lattice Fermions}

\author{\speaker{Emilie Huffman}%
         \thanks{Work done in collaboration with Shailesh Chandrasekharan and supported by DOE grant \#DEFG0205ER41368.}\\
        Duke University\\
        E-mail: \email{emilie.huffman@duke.edu}}

\abstract{We present a solution to the sign problem in a class of particle-hole symmetric Hamiltonian lattice fermion models on bipartite lattices using the idea of fermion bags. The solution remains valid when the particle-hole symmetry is broken through a staggered chemical potential term. This solution allows, for the first time, simulations of some massless four-fermion models with minimal fermion doubling and with an odd number of fermion flavors using ultra-local actions. One can thus study a variety of quantum phase transitions that have remained unexplored so far due to sign problems.}

\FullConference{The 32nd International Symposium on Lattice Field Theory\\
                 23-28 June, 2014\\
                 Columbia University New York, NY}

\begin{document}

\section{Introduction}

The theoretical study of strongly correlated quantum many-body systems is an exciting area of research with applications to a variety of physical systems that span many subfields of physics. However there still remain challenges to computing the observables of interest, especially for the systems that involve fermions. Calculating such obervables often involves sums with very large numbers of terms, and so a natural strategy would be to employ Monte Carlo methods to calculate the observables in a reasonable amount of time. However, the issue is that due to the quantum nature of these systems, the Monte Carlo weights may be negative or complex. In these cases, it is not clear how to assign weights to the configurations, and to formulate the problem in terms of positive weights is not a trivial task. When performed naively, the computational time scales as an exponential in the system's size, and this is referred to as a \textit{sign problem}. To find a representation such that positive weights are computable in polynomial time is called a solution to the sign problem, and there remain many unsolved sign problems in a variety of models of interest \cite{Chandrasekharan:2013rpa}.

These computational issues are often exacerbated if we wish to deal with strongly correlated fermions. Due to the Pauli principle, fermions are represented by operators or grassman numbers that anticommute with each other. Thus the weights of the different configurations are often both positive and negative, with little reason to favor the positive ones. These systems therefore often suffer severe sign problems \cite{Chandrasekharan:2002vk}.

The traditional approach for solving fermion sign problems is to first sum over all the fermionic degrees of freedom, leaving a sum of determinants involving a background potential. However, these terms still may not be positive, leaving plenty of models for which traditional methods do not work. A recently developed method, known as the fermion bag technique, involves partitioning fermionic degrees of freedom into several groups known as \textit{bags}, and summing over these bags. In a class of problems, which includes problems that are unsolvable with the traditional approach, fermion bags can be chosen such that their weights are positive \cite{Chandrasekharan:2013rpa}. The fermion bag approach then allows for a straightforward application of Monte Carlo techniques. Here we will apply this technique to a calculation involving staggered fermions.

The staggered fermion Hamiltonian is a discretized version of the Dirac Hamiltonian, which introduces a single fermion field component to each lattice site and interprets doubling as physical flavors. In the Lagrangian formalism on a space-time lattice, staggered fermions in $(2+1)$ dimensions describe two flavors of four-component Dirac fermions. However, it is difficult to maintain an $SU(2)$ flavor symmetry in the presence of interactions.

In contrast, a continuous time Hamiltonian formalism for staggered fermions describes a single flavor of four-component Dirac fermions because there is no doubling coming from time discretization. We are then free to add a second flavor of fermion to the staggered Hamiltonian, and we will get two flavors of Dirac fermions that respect an $SU(2)$ flavor symmetry. Thus Hamiltonian staggered fermions are useful for preserving flavor symmetries.

An additional advantage offered by the continuous time formalism is the restoration of particle-hole symmetry. Because the staggered fermion Hamiltonian is particle-hole symmetric, we should observe that the average occupation state, $\left\langle n\right\rangle$, is equal to $\frac{1}{2}$. However, calculations in the discrete time formalism show that this is not the case. Only in the limit as the time step $\epsilon \rightarrow 0$ do we have that $\left\langle n\right\rangle = \frac{1}{2}$. Thus the continuous time formalism Hamiltonian recovers this symmetry. We will see that this symmetry offers a pairing mechanism that is key in solving some sign problems.

\section{The Solution}

We now use the fermion bag technique to solve the sign problem for a single flavor of staggered fermions with a nearest neighbor Hubbard type interaction. While our example is in $(2+1)$ dimensions, the proof works in exactly the same way for higher dimensions. The ideas presented here are simple extensions of the solution we found in a non-relativistic spin polarized system recently \cite{Huffman:2013mla}. Since the sign problem is solved using the Hamiltonian formulation rather than the action formulation, continuous time calculations should be possible. The model is given by
\begin{equation}
H = t\sum_{x,y} c^\dagger_x M_{xy} c_y + \frac{V}{4} \sum_{\left\langle x,y\right\rangle} \left(n^+_x - n^-_x\right) \left(n^+_y - n^-_y\right),
\label{model1}
\end{equation}
where $n^+_x$ is the occupation number $c^\dagger_x c_x$, and $n^-_x = 1 - n^+_x$ is a hole operator. $M_{xy}$ is defined as:
\begin{equation}
M_{xy} = \frac{i}{2} \left(\delta_{x + \hat{\alpha}_1, y} - \delta_{x - \hat{\alpha}_1, y}\right) + \frac{i}{2} \left(-1\right)^{x^{1}} \left(\delta_{x + \hat{\alpha}_2, y} - \delta_{x - \hat{\alpha}_2, y}\right).
\end{equation}
Here $x$ and $y$ are two-dimensional spatial vectors, $\hat{\alpha}_1$ and $\hat{\alpha}_2$ are unit vectors in the two basis directions, and $x^1$ and $x^2$ refer to the first and second components of the $x$ vector, respectively. This Hamiltonian has a particle-hole symmetry, meaning it is unchanged under the transformations $c_x\rightarrow c_x^\dagger$ and $c_x^\dagger\rightarrow c_x$. A similar model known as the t-V model was considered a long time ago \cite{Sugar:1985}. In the repulsive case for $V \geq 2t$, the t-V model could be solved using a non-traditional method called the meron cluster approach \cite{Chandrasekharan:1999cm}, but that solution could not be extended to smaller values of $V$ where there is an interesting quantum phase transition.

If we were to employ the traditional Monte Carlo approach, we would start with the following expansion,
\begin{equation}
Z = Tr\left(e^{-\epsilon H} e^{-\epsilon H} ... e^{-\epsilon H}\right),
\end{equation}
where $\epsilon$ is small, and there are $N$ factors of $e^{-\epsilon H}$ so that $N \epsilon = \beta$. We would then write as a path integral:
\begin{equation}
\begin{array}{ccc}
	Z &=& \displaystyle \int \left[d\bar{\psi} d\psi \right] e^{-\bar{\psi}_1 \psi_1}\left\langle -\bar{\psi}_1 \right| e^{-\epsilon H} \left|\psi_2\right\rangle e^{-\bar{\psi}_2 \psi_2}\dots\\
&& \displaystyle \dots \left\langle \bar{\psi}_{N-1} \right| e^{-\epsilon H} \left|\psi_{N}\right\rangle e^{-\bar{\psi}_N \psi_N} \left\langle \bar{\psi}_N \right| e^{-\epsilon H} \left|\psi_1\right\rangle,
\end{array}
\end{equation}
\begin{equation}
 = \int \left[d\bar{\psi} d\psi \right] e^{-S\left[\bar{\psi}, \psi\right]}.
 \label{action}
\end{equation}
An auxiliary field method may be used to write the action so that it is quadratic and we could then write (\ref{action}) as a sum of determinants \cite{Hirsch}. However, there would be no guarantee of these determinants' positivity, and because particle-hole symmetry is lost, we would find sign problems in our model (as well as for many other models with an odd number of flavors).

Instead of the traditional method we will use a different approach, proceeding as follows. We refer to the first term in Eq. (\ref{model1}) as $H_0$ and write the interaction term, $H_{\rm int}$, more compactly as
\begin{equation}
H_{\rm int} = \frac{V}{4}\ \sum_{b,s_x,s_y} (s_x n^{s_x}_x) \ (s_y n^{s_y}_y).
\label{intchoice}
\end{equation}
Here the variable $b = \langle x,y\rangle$ labels the bond connecting the nearest neighbor sites $x$ and $y$, and $s_x,s_y = \pm 1$ label the presence of either $n^+$ or $-n^-$ at sites $x$ and $y$.

We may apply a unitary transformation to the staggered fermion Hamiltonian to obtain
\begin{equation}
M_{xy} = \frac{1}{2}\left(\delta_{x + \hat{\alpha}_1,y} + \delta_{x - \hat{\alpha}_1,y}\right) + \frac{\left(-1\right)^{x^1}}{2}\left(\delta_{x + \hat{\alpha}_2,y} + \delta_{x - \hat{\alpha}_2,y}\right).
\end{equation}
This real matrix $M$ satisfies this special property:
\begin{equation}
M^T = -D M D,
\label{symm}
\end{equation}
where $D_{xy} = \sigma_x \delta_{xy}$ is a diagonal matrix with elements $\sigma_x$, and $\sigma_x = \left(-1\right)^{x^1 + x^2}$. Thus we have $\sigma_x = +1$ for $x$ belonging to an even site and $\sigma_x = -1$ for $x$ belonging to an odd site. This property of $M$ will play an important role in the solution to the sign problem.

We then use the well known series expansion of the partition function that is used often for continuous time Monte Carlo methods \cite{PhysRevLett.81.2514,PhysRevB.72.035122,PhysRevE.74.036701,PhysRevLett.101.090402,PhysRevA.82.053621,RevModPhys.83.349}. 
For our model, we obtain the following:
\begin{equation}
Z = \sum_k \sum_{[b,s]} \int [dt] \big(-V/4\big)^k \mathrm{Tr}\Big(\mathrm{e}^{-(\beta-t_1) H_0} (s_1 n^{s_1}_{x_1}) \mathrm{e}^{-(t_2-t_1)H_0} (s_2 n^{s_2}_{x_2}) \dots \mathrm{e}^{-(t_{2k-1}-t_{2k}) H_0} (s_{2k} n^{s_{2k}}_{x_{2k}}) \mathrm{e}^{-t_k H_0} \Big),
\label{sse}
\end{equation}
where $[b,s]$ defines a configuration of $k$ interaction bonds located at times $t_1 \geq t_2 \geq t_3\geq...\geq t_{k}$, and $[dt]$ represents the $k$ time-ordered integrations from $0$ to $\beta$ over these locations of the interaction bonds. Each of the $k$ bond interactions contains two interaction vertices. We label these $2k$ interaction vertices with the index $q=1,2...,2k$ such that $x_q$ labels the spatial site of the vertex, $t_q$ labels the temporal location of the vertex, and $s_q$ labels the particle-hole operator that is inserted at the vertex. Furthermore, since the interactions are bonds we will naturally have $t_1 = t_2 \geq t_3 = t_4 \geq ...\geq t_{2k-1}= t_{2k}$. Thus the integration $[dt]$ involves only $k$ integrations.

Using standard manipulations in the Fock space formalism of quantum many body theory, we can evaluate (\ref{sse}) exactly in terms of a determinant of a $2k \times 2k$ matrix $G([b,s,t])$, allowing the following expression for Z:
\begin{equation}
Z \ = \ Z_0 \sum_k \sum_{[b,s]}\ \int\ [dt]\ (-V/4)^k \  \mathrm{Det} G([b,s,t]),
\label{sse1}
\end{equation}
where $Z_0$ is the free partition function. Note that the expression is written in continuous time and so may be used in continuous time Monte Carlo. The matrix elements of $G([b,s,t])$ can be obtained from the two point functions
\begin{eqnarray}
\mathrm{Tr}\Big(\mathrm{e}^{-(\beta-t)H_0} c_{x_q}  \mathrm{e}^{-t H_0} c^\dagger_{x_{q'}}\Big) &=& \Big(\frac{\mathrm{e}^{-tM}}{1 + \mathrm{e}^{-\beta M}}\Big)_{x_q,x_{q'}} \\
\mathrm{Tr}\Big(\mathrm{e}^{-(\beta-t)H_0} c^\dagger_{x_q}  \mathrm{e}^{-t H_0} c_{x_{q'}}\Big) &=& \Big(\frac{\mathrm{e}^{t M^T}}{1 + \mathrm{e}^{\beta M^T}}\Big)_{x_q,x_{q'}}
\end{eqnarray}
using Wick's theorem. The properties of (\ref{symm}) can then be used along with known results to show
\begin{equation}
\Big(\frac{\mathrm{e}^{t M^T}}{1 + \mathrm{e}^{\beta M^T}}\Big)_{x_q,x_{q'}} = \ \ 
\sigma_{x_{q}} \sigma_{x_{q'}} \Big(\frac{\mathrm{e}^{-t M}}{1 + \mathrm{e}^{-\beta M}}\Big)_{x_{q},x_{q'}}.
\end{equation}
Using these results we have shown that for a fixed $q < q'$, the off-diagonal matrix elements of $G([b,s,t])$ are given by the following formulae:
\begin{eqnarray}
G_{qq'}([b,s,t]) \ &=&\ \Big(\frac{\mathrm{e}^{-(t_q-t_{q'}) M}}{1 + \mathrm{e}^{-\beta M}}\Big)_{x_{q},x_{q'}} 
\label{offdiag1}
\\
G_{q'q}([b,s,t]) \ &=&\ -\ \sigma_{x_{q}} \ \sigma_{x_{q'}}\ G_{qq'}([b,s,t])
\label{offdiag2}
\end{eqnarray}
which do not depend on $[s]$. The negative sign in (\ref{offdiag2}) is due to the usual anti-periodic boundary conditions in the time dimension, which must be introduced when the trace is written as a determinant. For $q = q'$, the diagonal matrix elements are given by 
\begin{equation}
G_{qq'}([b,s,t]) \ =\ -\frac{s_q}{2}\ \delta_{q q'},
\label{diag}
\end{equation} 
and depend only on $[s]$. It is interesting to note that $[s]$ does not enter the formulae for the off-diagonal matrix elements (\ref{offdiag1}) and (\ref{offdiag2}), which at first seems surprising and in fact will play a large role in our solution. However, note that the insertion of $s_q n^{s_q}_{x_q}$ implies either the operator $c^\dagger_{x_{q}} c_{x_{q}}$ or the operator $-c_{x_{q}} c^\dagger_{x_{q}}$, and by anticommutation relations, $c^\dagger_{x_{q}} c_{x_{q}} = -c_{x_{q}} c^\dagger_{x_{q}} + 1$, meaning both are the same operator except for diagonal terms. Hence $[s]$ only enters through diagonal terms.

\begin{figure}  
\begin{center}  
\includegraphics[width=14cm]{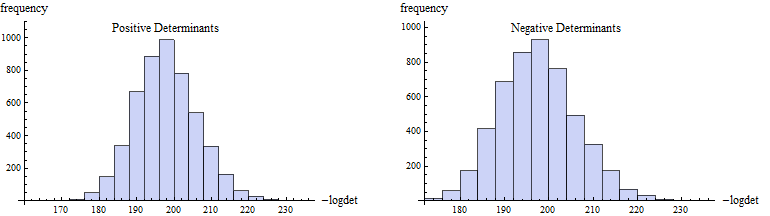}  
\caption{In a sample of 10,000 randomly generated configurations $\left[b, s, t\right]$ on an $8\times8$ square lattice at $\beta = 10$ involving the matrices of size $250\times250$, we obtained 5004 positive determinants and 4996 negative determinants. Histograms of positive determinants (left) and negative determinants (right) in this sample are plotted above. The similarity of the two distributions suggests the existence of a severe sign problem.}
\label{sign}
\end{center}  
\end{figure}

We note that $\left[s\right]$ as a single configuration does not respect particle-hole symmetry, as $\left[s\right] \rightarrow \left[-s\right]$, under the transformation. Thus we are missing that possibility of a pairing mechanism and when we calculate the determinants of randomly selected $G\left[b,s,t\right]$ matrices, we find the severe sign problem shown in figure \ref{sign}, with as many negative determinants as positive determinants.

Here is where the fermion bag approach comes into play. As explained in \cite{Chandrasekharan:2013rpa}, diagonal elements of the fermion matrix can be treated as fermion bags. Since the variables $[s]$ appear only through diagonal terms, a sum over all possible $[s]$ configurations should be possible by treating these diagonal terms as fermion bags. We can then show how the technique works mathematically by examining our expression using the Grassmann integral form of a determinant. Here we write $G([b,s,t]) = D_0([s]) + A([b,t])$, where $D_0([s])$ is the diagonal part defined in (\ref{diag}), and $A([b,t])$ is the off-diagonal part defined in (\ref{offdiag1}) and (\ref{offdiag2}). Using Grassmann integrals, we have
\begin{eqnarray}
\sum_{[s]} \mathrm{Det}(G[b,s,t]) &=& \sum_{[s]} \int [d\bar{\psi} \ d\psi] \mathrm{e}^{-\bar{\psi} (D_0([s]) + A([b,t]))\psi} . \nonumber \\
\end{eqnarray}
Substituting $D_0([s])$ from (\ref{diag}) and performing the $[s]$ sum first, we obtain
\begin{equation}
\sum_{[s]} \ \mathrm{e}^{-\bar{\psi} D_0([s])\psi} = \prod_q \sum_{s_q=1,-1} (1 - \frac{s_q}{2} \bar{\psi}_q \psi_q) = 4^k.
\end{equation}
Thus $\sum_{[s]} \mathrm{Det}(G[b,s,t]) = 4^k \mathrm{Det} (A([b,t]))$, and the diagonal terms disappear from $Z$. Substituting this result into (\ref{sse1}) we obtain
\begin{equation}
Z \ = \ \sum_{[b]}\ \int\ [dt]\ (-V)^k \ \mathrm{Det} (A([b,t])).
\label{sse2}
\end{equation}
We now have a sum of fermion bag terms, and can show that these fermion bag terms are all positive.

The matrix $A([b,t])$, defined in (\ref{offdiag1}) and (\ref{offdiag2}), is real and satisfies the relation $A^T = - \tilde{D} A \tilde{D}$, where $\tilde{D}$ is the diagonal matrix obtained from $D$ but restricted to the $2k$ interaction sites (i.e., $\tilde{D}_{x_q,x_{q'}} = \sigma_{x_q}\ \delta_{x_q,x_{q'}}$). Thus $A\tilde{D}$ is a real anti-symmetric matrix whose determinant must be positive.  But $\mathrm{Det}(\tilde{D}) = (-1)^k$ because half the sites are even, and half are odd. Thus,
\begin{equation}
(-1)^k\mathrm{Det}(A([b,t]))\ =\ \mathrm{Det}(A\tilde{D})  \geq 0,
\end{equation}
and we finally obtain
\begin{equation}
Z \ = \ \sum_{[b]}\ \int\ [dt]\ V^k \  \mathrm{Det} (A([b,t])\tilde{D})
\end{equation}
which is a sum over positive terms for $V > 0$. The sign problem for this tight binding model has thus been solved using the fermion bag method and the formulae (\ref{offdiag1}), (\ref{offdiag2}), and (\ref{diag}).

\section{Conclusion}

From this work we have seen that even in models that include particle-hole symmetry, there can still be sign problems if traditional methods are used. Particle-hole symmetry is lost when the discrete time traditional method is used. However, it is recovered in the continuous time Hamiltonian formalism and we have shown that we are able to solve a class of particle-hole symmetric models in this formalism using the fermion bag technique: specifically models where the free Hamiltonian satisfies (\ref{symm}), and interaction vertices respect a \textit{staggered reference configuration}, consisting of particles on the even sublattice and holes on the odd sublattice. Violations to this configuration may be introduced, so long as they are introduced in a correlated fashion such that a controlled resummation over positive and negative configurations may be performed. For example, a staggered chemical potential term
\begin{equation}
H_{\rm stagg} = \sum_i h_i n_i^{s_i},
\end{equation}
where $h_i \geq 0$, and $s_i$ is $+1$ on the even sublattice and $-1$ on the odd sublattice may be introduced without causing sign problems. In addition to the four-fermion models and t-V model, other models with an odd number of flavors, such as $SU(3)$ Gross-Neveu models, may be solved. More details about solvable models are given in \cite{Huffman:2013mla}. With this method it is now possible to study new quantum critical behavior using quantum Monte Carlo methods, including graphene-inspired systems on a honeycomb lattice \cite{Wang:2014cbw}, staggered fermions, and three-flavor nuclear inspired systems \cite{Huffman:2013mla}.


\begin{thebibliography}{99}
\bibitem{Chandrasekharan:2013rpa} 
  S.~Chandrasekharan,
  Eur.\ Phys.\ J.\ A {\bf 49}, 90 (2013)
  [arXiv:1304.4900 [hep-lat]].
 \bibitem{Chandrasekharan:2002vk} 
  S.~Chandrasekharan, J.~Cox, J.~C.~Osborn and U.~J.~Wiese,
  Nucl.\ Phys.\ B {\bf 673}, 405 (2003)
  [cond-mat/0201360].
\bibitem{Huffman:2013mla} 
  E.~F.~Huffman and S.~Chandrasekharan,
  Phys.\ Rev.\ B {\bf 89}, 111101 (2014)
  [arXiv:1311.0034 [cond-mat.str-el]].
\bibitem{Sugar:1985}
  J.~E.~Gubernatis, D.~J.~Scalapino, R.~L.~Sugar, and W.~D.~Toussaint,
  Phys.\ Rev.\ B {\bf 32}, 103 (1985).
\bibitem{Chandrasekharan:1999cm} 
  S.~Chandrasekharan and U.~J.~Wiese,
  Phys.\ Rev.\ Lett.\  {\bf 83}, 3116 (1999)
  [cond-mat/9902128].
\bibitem{Hirsch}
	J.~E.~Hirsch,
	Phys.\ Rev. B {\bf 28} 7 (1983).
\bibitem{PhysRevLett.81.2514}
	N.V.~Prokof'ev and B.V.~Svistunov,
	Phys.\ Rev.\ Lett.\ {\bf 81}, 12 (1998),
	[arXiv:cond-mat/9804097v1].
\bibitem{PhysRevB.72.035122}
	A.N.~Rubtsov, V.V.~Savkin, and A.I.~Lichtenstein,
	Phys.\ Rev.\ B.\ {\bf 72}, 3 (2005),
	[arXiv:cond-mat/0411344v2].
\bibitem{PhysRevE.74.036701}
	M.~Boninsegni, N.V.~Prokof'ev, and B.V.~Svistunov,
	Phys.\ Rev.\ E\ {\bf 74}, 3, 2006,
	[arXiv:physics/0605225v2 [physics.comp-ph]].
\bibitem{PhysRevLett.101.090402}
	E.~Burovski, E.~Kozik, N.~Prokof'ev, B.~Svistunov, and M.~Troyer,
	Phys.\ Rev.\ Lett.\ {\bf 101}, 9, (2008),
	[arXiv:0805.3047v2 [cond-mat.str-el]].
\bibitem{PhysRevA.82.053621}
	O.~Goulko, and M.~Wingate,
	Phys.\ Rev.\ A.\ {\bf 82}, 5, (2010),
	[arXiv:1008.3348v2 [cond-mat.quant-gas]].
\bibitem{RevModPhys.83.349}
	E.~Gull, A.J.~Millis, A.I.~Lichtenstein, and A.N.~Rubtsov,
	Rev.\ Mod.\ Phys.\ {\bf 83}, 2, (2011),
	[arXiv:1012.4474v2 [cond-mat.str-el]].
\bibitem{Wang:2014cbw} 
  L.~Wang, P.~Corboz and M.~Troyer,
  New J.\ Phys.\  {\bf 16}, no. 10, 103008 (2014)
  [arXiv:1407.0029 [cond-mat.str-el]].
\end{thebibliography}
\end{document}